\documentclass[numbers,sort&compress]{IntechOpen-Book}
\usepackage{amsmath,amsthm}
\usepackage{amsfonts}
\usepackage{amssymb}
\usepackage{amsbsy,bm}
\renewcommand{\mathbf}{\bm}
\renewcommand{\mathsf}{\hat}
\renewcommand{\mathtt}{\mathrm}

\newcommand{\changeurlcolor}[1]{\hypersetup{urlcolor=blue}}

\begin{document}

\Mainmatter

\begin{frontmatter}

\chapter{Obscure qubits and membership amplitudes}
\author{Steven Duplij and
Raimund Vogl\\
Center for Information Technology (WWU IT),
University of M\"unster,
D-48149 M\"unster,
Deutschland}

\makechaptertitle

\chaptermark{Obscure qubits and membership amplitudes}

\begin{abstract}
We propose a concept of quantum computing which incorporates an additional kind of uncertainty, i.e. vagueness (fuzziness), in a natural way by introducing new entities, obscure qudits (e.g. obscure qubits), which are  characterized simultaneously by a quantum probability and by a membership function. To achieve this, a
membership amplitude for quantum states is introduced alongside the quantum amplitude. The Born rule is used for the
quantum probability only, while the membership function can be computed
from the membership amplitudes according to a chosen model. Two different
versions of this approach  are given here: the \textquotedblleft product\textquotedblright\ obscure
qubit, where the resulting amplitude is a product of the quantum amplitude
and the membership amplitude, and the \textquotedblleft
Kronecker\textquotedblright\ obscure qubit, where quantum and vagueness
computations are to be performed independently (i.e. quantum computation alongside truth evaluation).
The latter is called a double obscure-quantum computation.
In this case, the measurement becomes mixed in the quantum and obscure amplitudes,
while the density matrix is not idempotent. The obscure-quantum gates act not in the tensor
product of spaces, but in the direct product of quantum Hilbert space and so called membership
space which are of different natures and properties.
The concept of double (obscure-quantum) entanglement is introduced, and vector and scalar
concurrences are proposed, with some examples being given.
\end{abstract}

\begin{keywords}
qubit, fuzzy, membership function, amplitude, Hilbert space
\end{keywords}

\end{frontmatter}

\section{Introduction}

Nowadays, the development of quantum computing technique is governed by theoretical  
extensions of its ground concepts \cite{nie/chu,kay/laf/mos,wil/cle}. One of
them is to allow two kinds of uncertainty, sometimes called
randomness and vagueness/fuzziness (for a review, see, \cite{goo/ngu}), which
leads to the formulation of combined probability and possibility theories
\cite{dub/ngu/pra} (see, also, \cite{belohl,dub/pra,smith,zim11}). Various
interconnections between vagueness and quantum probability calculus were
considered in \cite{pykacz,dvu/cho,bar/rie/tir,gra94}, including the treatment of
inaccuracy in measurements \cite{gud88,gud2005}, non-sharp amplitude
densities \cite{gud89} and the related concept of partial Hilbert spaces \cite{gud86}.

Relations between truth values and probabilities were also given in \cite{bol18}. The
hardware realization of computations with vagueness was considered in
\cite{hir/oza89,virant}. On the fundamental physics side, it was shown that the
discretization of space-time at small distances can lead to a discrete (or
fuzzy) character for the quantum states themselves
\cite{bun/hsu/zee2005,mar/ruz2012,bar2007,mue2009}.

With a view to applications of the above ideas in quantum
computing, we introduce a definition of quantum state which is described by both a
quantum probability and a membership function, and thereby
incorporate vagueness/fuzziness directly into the formalism. In
addition to the probability amplitude we will define a membership amplitude, and
such a state will be called an obscure/fuzzy qubit (or qudit).

In general, the Born rule will apply to the quantum probability alone,
while the membership function can be taken to be an arbitrary function of all the amplitudes
fixed by the chosen model of vagueness. Two different models of
\textquotedblleft obscure-quantum computations with truth\textquotedblright are proposed below :
1) A \textquotedblleft Product\textquotedblright\ obscure qubit, in which the
resulting amplitude is the product (in $\mathbb{C}$) of the quantum amplitude
and the membership amplitude; 2) A \textquotedblleft Kronecker\textquotedblright%
\ obscure qubit for which computations are performed \textquotedblleft in
parallel\textquotedblright, so that quantum amplitudes and the membership
amplitudes form \textquotedblleft vectors\textquotedblright, which we will call
obscure-quantum amplitudes.
In the latter case, which we call a double obscure-quantum computation, the protocol of measurement
depends on both the quantum and obscure amplitudes, and in this case the density matrix need not be idempotent.
We define a new kind of ``gate'', namely, the obscure-quantum gates, which are linear transformations
in the direct product (not in the tensor product) of spaces: a quantum Hilbert space and
a so-called membership space having special fuzzy properties.
We introduce a new concept of double (obscure-quantum) entanglement, in which vector and scalar
concurrences are defined and computed for some examples.

\section{Preliminaries}

To establish notation standard in the literature (see, e.g. \cite{nie/chu,kay/laf/mos} and
\cite{bry/bry,run/mun,kim2003}) we present the following definitions. In an underlying $d$-dimensional Hilbert space, the standard qudit (using the computational
basis and Dirac notation)
$\mathcal{H}_{q}^{\left(  d\right)  }$ is given by

\begin{equation}
\left\vert \psi^{\left(  d\right)  }\right\rangle =\sum_{i=0}^{d-1}%
a_{i}\left\vert i\right\rangle ,\ \ \ \ a_{i}\in\mathbb{C},\left\vert
i\right\rangle \in\mathcal{H}_{q}^{\left(  d\right)  }, \label{pn}%
\end{equation}
where $a_{i}$ is a probability amplitude of the state $\left\vert i\right\rangle$. (For a review, see, e.g.
\cite{gen/tra,wan/hu/san/kai})%
 The probability $p_{i}$ to measure
the $i$th state is $p_{i}=F_{p_{i}}\left(  a_{1},\ldots,a_{n}\right)  $,
$0\leq p_{i}\leq1$, $0\leq i\leq d-1$. The shape of the functions $F_{p_{i}}$
is governed by the Born rule $F_{p_{i}}\left(  a_{1},\ldots,a_{d}\right)
=\left\vert a_{i}\right\vert ^{2}$, and $\sum_{i=0}^{d}p_{i}=1$. A one-qudit
($L=1$) quantum gate is a unitary transformation $U^{\left(  d\right)
}:\mathcal{H}_{q}^{\left(  d\right)  }\rightarrow\mathcal{H}_{q}^{\left(
d\right)  }$ described by unitary $d\times d$ complex matrices acting on the
vector (\ref{pn}), and for a register containing $L$ qudits quantum gates are
unitary $d^{L}\times d^{L}$ matrices.
The quantum circuit model \cite{deu2,bar/ben/cle} forms the basis for the standard concept of quantum computing. Here the quantum algorithms are compiled as a sequence of elementary gates acting on a register containing $L$ qubits (or qudits), followed by a measurement to yield the result \cite{llo95,bry/bry}.

For further details on qudits and their transformations, see for example the reviews \cite{gen/tra,wan/hu/san/kai} and the references therein.

\section{Membership amplitudes}
We define an obscure qudit with $d$
states via the following superposition (in place of that given in (\ref{pn}))%

\begin{equation}
\left\vert \psi_{ob}^{\left(  d\right)  }\right\rangle =\sum_{i=1}^{d-1}%
\alpha_{i}a_{i}\left\vert i\right\rangle , \label{pob}%
\end{equation}
where $a_{i}$ is a (complex) probability amplitude $a_{i}\in\mathbb{C}$, and we have introduced a (real) membership amplitude
$\alpha_{i}$ , with $\alpha_{i}\in\left[
0,1\right]  $, $0\leq i\leq d-1$. The probability $p_{i}$ to
find the $i$th
state upon measurement, and the membership function $\mu_{i}$ (\textquotedblleft of
truth\textquotedblright)for the $i$th state are both functions of the corresponding
amplitudes as follows%
\begin{align}
p_{i}  &  =F_{p_{i}}\left(  a_{0},\ldots,a_{d-1}\right)  ,\ \ \ 0\leq
p_{i}\leq1,\label{p}\\
\mu_{i}  &  =F_{\mu_{i}}\left(  \alpha_{0},\ldots,\alpha_{d-1}\right)
,\ \ \ 0\leq\mu_{i}\leq1. \label{m}%
\end{align}

The dependence of the probabilities of the $i$th states upon the amplitudes,
i.e. the form of the function $F_{p_{i}}$ is fixed by the Born rule%
\begin{equation}
F_{p_{i}}\left(  a_{1},\ldots,a_{n}\right)  =\left\vert a_{i}\right\vert ^{2},
\label{b}%
\end{equation}
while the form of $F_{\mu_{i}}$ will vary according to different
obscurity assumptions. In this paper we consider only real membership amplitudes
and membership functions (complex obscure sets and numbers
were considered in \cite{buc89,ram/mil/fri/kan,gar2012}). In this context  the real
functions $F_{p_{i}}$ and $F_{\mu_{i}}$, $0\leq i\leq d-1$ will contain
complete information about the obscure qudit (\ref{pob}).

We impose the normalization conditions%
\begin{align}
\sum_{i=0}^{d-1}p_{i}  &  =1,\label{p1}\\
\sum_{i=0}^{d-1}\mu_{i}  &  =1, \label{m1}%
\end{align}
where the first condition is standard in quantum mechanics, while the second
condition is taken to hold by analogy. Although (\ref{m1}) may not be satisfied,
we will not consider that case.

For $d=2$, we obtain for the obscure qubit the general form (instead of that in
(\ref{pob}))%

\begin{align}
&  \left\vert \psi_{ob}^{\left(  2\right)  }\right\rangle =\alpha_{0}%
a_{0}\left\vert 0\right\rangle +\alpha_{1}a_{1}\left\vert 1\right\rangle
,\label{p2}\\
&  F_{p_{0}}\left(  a_{0},a_{1}\right)  +F_{p_{1}}\left(  a_{0},a_{1}\right)
=1,\\
&  F_{\mu_{0}}\left(  \alpha_{0},\alpha_{1}\right)  +F_{\mu_{1}}\left(
\alpha_{0},\alpha_{1}\right)  =1.
\end{align}

The Born probabilities to observe the states $\left\vert 0\right\rangle $ and
$\left\vert 1\right\rangle $ are%
\begin{equation}
p_{0}=F_{p_{0}}^{Born}\left(  a_{0},a_{1}\right)  =\left\vert a_{0}\right\vert
^{2},\ \ \ p_{1}=F_{p_{1}}^{Born}\left(  a_{0},a_{1}\right)  =\left\vert
a_{1}\right\vert ^{2}, \label{fb}%
\end{equation}
and the membership functions are%
\begin{equation}
\mu_{0}=F_{\mu_{0}}\left(  \alpha_{0},\alpha_{1}\right)  ,\ \ \ \mu_{1}%
=F_{\mu_{1}}\left(  \alpha_{0},\alpha_{1}\right)  . \label{mm}%
\end{equation}
If we assume the Born rule (\ref{fb}) for the membership functions as well%
\begin{equation}
F_{\mu_{0}}\left(  \alpha_{0},\alpha_{1}\right)  =\alpha_{0}^{2}%
,\ \ \ \ F_{\mu_{1}}\left(  \alpha_{0},\alpha_{1}\right)  =\alpha_{1}^{2},
\label{mm1}%
\end{equation}
(which is one of various possibilities depending on the chosen model), then%
\begin{align}
\left\vert a_{0}\right\vert ^{2}+\left\vert a_{1}\right\vert ^{2}  &
=1,\label{n1}\\
\alpha_{0}^{2}+\alpha_{1}^{2}  &  =1. \label{n2}%
\end{align}

Using (\ref{n1})--(\ref{n2}) we can parametrize (\ref{p2}) as%
\begin{align}
\left\vert \psi_{ob}^{\left(  2\right)  }\right\rangle  &  =\cos\frac{\theta
}{2}\cos\frac{\theta_{\mu}}{2}\left\vert 0\right\rangle +e^{i\varphi}\sin
\frac{\theta}{2}\sin\frac{\theta_{\mu}}{2}\left\vert 1\right\rangle ,\\
0  &  \leq\theta\leq\pi,\ \ \ \ 0\leq\varphi\leq2\pi,\ \ \ \ 0\leq\theta_{\mu
}\leq\pi.
\end{align}
Therefore, obscure qubits (with Born-like rule for the
membership functions) are geometrically described by
a pair of vectors, each inside a Bloch ball (and not as vectors on the boundary spheres,
because \textquotedblleft$\left\vert \sin\right\vert
,\left\vert \cos\right\vert \leq1$\textquotedblright), where one is for the
probability amplitude (an ellipsoid inside the Bloch ball with $\theta_{\mu
}=const_{1}$), and the other for the membership amplitude (which is reduced
to an ellipse, being a slice inside the Bloch ball with $\theta=const_{2}$,
$\varphi=const_{3}$). The norm of the obscure qubits is not constant however, because%
\begin{equation}
\left\langle \psi_{ob}^{\left(  2\right)  }\mid\psi_{ob}^{\left(  2\right)
}\right\rangle =\frac{1}{2}+\frac{1}{4}\cos\left(  \theta+\theta_{\mu}\right)
+\frac{1}{4}\cos\left(  \theta-\theta_{\mu}\right)  . \label{n}%
\end{equation}
In the case where $\theta=\theta_{\mu}$, the norm (\ref{n}) becomes $1-\frac{1}%
{2}\sin^{2}\theta$, reaching its minimum $\frac{1}{2}$ when $\theta
=\theta_{\mu}=\frac{\pi}{2}$.

Note that for complicated functions $F_{\mu_{0,1}}\left(  \alpha_{0}%
,\alpha_{1}\right)  $ the condition (\ref{n2}) may be not satisfied, but the
condition (\ref{m1}) should nevertheless always be valid. The concrete form of
the functions $F_{\mu_{0,1}}\left(  \alpha_{0},\alpha_{1}\right)  $ depends
upon the chosen model. In the simplest case, we can identify two arcs on the
Bloch ellipse for $\alpha_{0},\alpha_{1}$ with the membership functions and
obtain%
\begin{align}
F_{\mu_{0}}\left(  \alpha_{0},\alpha_{1}\right)   &  =\frac{2}{\pi}%
\arctan\frac{\alpha_{1}}{\alpha_{0}},\label{f1}\\
F_{\mu_{1}}\left(  \alpha_{0},\alpha_{1}\right)   &  =\frac{2}{\pi}%
\arctan\frac{\alpha_{0}}{\alpha_{1}}, \label{f2}%
\end{align}
such that $\mu_{0}+\mu_{1}=1$, as in (\ref{m1}).

In \cite{man06,mar/vis/rei13} a two stage special construction of quantum
obscure/fuzzy sets was considered. The so-called classical-quantum
obscure/fuzzy registers were introduced in the first step (for $n=2$, the minimal case) as%
\begin{align}
\left\vert s\right\rangle _{f} &  =\sqrt{1-f}\left\vert 0\right\rangle
+\sqrt{f}\left\vert 1\right\rangle ,\\
\left\vert s\right\rangle _{g} &  =\sqrt{1-g}\left\vert 0\right\rangle
+\sqrt{g}\left\vert 1\right\rangle ,
\end{align}
where $f,g\in\left[  0,1\right]  $ are the relevant classical-quantum membership
functions. In the second step their quantum superposition is defined by%
\begin{equation}
\left\vert s\right\rangle =c_{f}\left\vert s\right\rangle _{f}+c_{g}\left\vert
s\right\rangle _{g},\label{s}%
\end{equation}
where $c_{f}$ and $c_{g}$ are the probability amplitudes of the fuzzy states
$\left\vert s\right\rangle _{f}$ and $\left\vert s\right\rangle _{g}$,
respectively. It can be seen that the state (\ref{s}) is a particular case of
(\ref{p2}) with%
\begin{align}
\alpha_{0}a_{0} &  =c_{f}\sqrt{1-f}+c_{g}\sqrt{1-g},\\
\alpha_{1}a_{1} &  =c_{f}\sqrt{f}+c_{g}\sqrt{g}.
\end{align}

This gives explicit connection of our double amplitude description of obscure
qubits with the approach \cite{man06,mar/vis/rei13} which uses probability
amplitudes and the membership functions. It is important to note that the use
of the membership amplitudes introduced here $\alpha_{i}$ and (\ref{pob})
allows us to exploit the standard quantum gates, but not to define new special
ones, as in \cite{man06,mar/vis/rei13}.

Another possible form of $F_{\mu_{0,1}}\left(  \alpha_{0},\alpha_{1}\right)  $
(\ref{mm}), with the corresponding membership functions satisfying the
standard fuzziness rules, can be found using a standard homeomorphism between the
circle and the square. In \cite{han/hat/hir,ryb/kag/rap} this transformation
was applied to the probability amplitudes $a_{0,1}$, but here we exploit it
for the membership amplitudes $\alpha_{0,1}$%
\begin{align}
F_{\mu_{0}}\left(  \alpha_{0},\alpha_{1}\right)   &  =\frac{2}{\pi}%
\arcsin\sqrt{\frac{\alpha_{0}^{2}\operatorname*{sign}\alpha_{0}-\alpha_{1}%
^{2}\operatorname*{sign}\alpha_{1}+1}{2}},\\
F_{\mu_{1}}\left(  \alpha_{0},\alpha_{1}\right)   &  =\frac{2}{\pi}%
\arcsin\sqrt{\frac{\alpha_{0}^{2}\operatorname*{sign}\alpha_{0}+\alpha_{1}%
^{2}\operatorname*{sign}\alpha_{1}+1}{2}}.
\end{align}

So for positive $\alpha_{0,1}$ we obtain (cf. \cite{han/hat/hir})%
\begin{align}
F_{\mu_{0}}\left(  \alpha_{0},\alpha_{1}\right)   &  =\frac{2}{\pi}%
\arcsin\sqrt{\frac{\alpha_{0}^{2}-\alpha_{1}^{2}+1}{2}},\\
F_{\mu_{1}}\left(  \alpha_{0},\alpha_{1}\right)   &  =1.
\end{align}

The equivalent membership functions for the outcome are%
\begin{align}
&  \max\left(  \min\left(  F_{\mu_{0}}\left(  \alpha_{0},\alpha_{1}\right)
,1-F_{\mu_{1}}\left(  \alpha_{0},\alpha_{1}\right)  \right)  ,\min\left(
1-F_{\mu_{0}}\left(  \alpha_{0},\alpha_{1}\right)  \right)  ,F_{\mu_{1}%
}\left(  \alpha_{0},\alpha_{1}\right)  \right)  ,\\
&  \min\left(  \max\left(  F_{\mu_{0}}\left(  \alpha_{0},\alpha_{1}\right)
,1-F_{\mu_{1}}\left(  \alpha_{0},\alpha_{1}\right)  \right)  ,\max\left(
1-F_{\mu_{0}}\left(  \alpha_{0},\alpha_{1}\right)  \right)  ,F_{\mu_{1}%
}\left(  \alpha_{0},\alpha_{1}\right)  \right)  .
\end{align}

There are many different models for $F_{\mu_{0,1}}\left(  \alpha_{0}%
,\alpha_{1}\right)  $ which can be introduced in such a way that they satisfy
the obscure set axioms \cite{dub/pra,zim11}.

\section{Transformations of obscure qubits}

Let us consider the obscure qubits in the vector representation, such that%
\begin{equation}
\left\vert 0\right\rangle =\left(
\begin{array}
[c]{c}%
1\\
0
\end{array}
\right)  ,\ \ \ \ \left\vert 1\right\rangle =\left(
\begin{array}
[c]{c}%
0\\
1
\end{array}
\right)  \label{vr}%
\end{equation}
are basis vectors of $\mathcal{H}_{q}^{\left(  2\right)  }$. Then a standard
quantum computational process in the quantum register with $L$ obscure qubits
(qudits (\ref{pn})) is performed by sequences of unitary matrices $\mathsf{U}$
of size $2^{L}\times2^{L}$ ($n^{L}\times n^{L}$), $\mathsf{U}^{\dag}%
\mathsf{U}=\mathsf{I}$, which are called quantum gates ($\mathsf{I}$ is the unit matrix).
Thus, for one obscure
qubit the quantum gates are $2\times2$ unitary complex matrices.

In the vector representation, an obscure qubit differs from the standard qubit
(\ref{p2}) by a $2\times2$ invertible diagonal (not necessarily unitary) matrix%
\begin{align}
\left\vert \psi_{ob}^{\left(  2\right)  }\right\rangle  &  =\mathsf{M}\left(
\alpha_{0},\alpha_{1}\right)  \left\vert \psi^{\left(  2\right)
}\right\rangle ,\label{ps}\\
\mathsf{M}\left(  \alpha_{0},\alpha_{1}\right)   &  =\left(
\begin{array}
[c]{cc}%
\alpha_{0} & 0\\
0 & \alpha_{1}%
\end{array}
\right)  . \label{ma}%
\end{align}
We call $\mathsf{M}\left(  \alpha_{0},\alpha_{1}\right)  $ a membership matrix
which can optionally have the property%
\begin{equation}
\operatorname*{tr}\mathsf{M}^{2}=1, \label{tm}%
\end{equation}
if (\ref{n2}) holds.

Let us introduce the orthogonal commuting projection operators%
\begin{align}
\mathsf{P}_{0}  &  =\left(
\begin{array}
[c]{cc}%
1 & 0\\
0 & 0
\end{array}
\right)  ,\ \ \ \ \mathsf{P}_{1}=\left(
\begin{array}
[c]{cc}%
0 & 0\\
0 & 1
\end{array}
\right)  ,\label{pp}\\
\mathsf{P}_{0}^{2}  &  =\mathsf{P}_{0},\ \ \mathsf{P}_{1}^{2}=\mathsf{P}%
_{1},\ \ \mathsf{P}_{0}\mathsf{P}_{1}=\mathsf{P}_{1}\mathsf{P}_{0}=\mathsf{0},
\label{pp1}%
\end{align}
where $\mathsf{0}$ is the $2\times2$ zero matrix. Well-known properties of the
projections are that%
\begin{align}
\mathsf{P}_{0}\left\vert \psi^{\left(  2\right)  }\right\rangle  &
=a_{0}\left\vert 0\right\rangle ,\ \ \ \ \ \mathsf{P}_{1}\left\vert
\psi^{\left(  2\right)  }\right\rangle =a_{1}\left\vert 0\right\rangle
,\label{ppp}\\
\left\langle \psi^{\left(  2\right)  }\right\vert \mathsf{P}_{0}\left\vert
\psi^{\left(  2\right)  }\right\rangle  &  =\left\vert a_{0}\right\vert
^{2},\ \ \ \ \ \left\langle \psi^{\left(  2\right)  }\right\vert
\mathsf{P}_{1}\left\vert \psi^{\left(  2\right)  }\right\rangle =\left\vert
a_{1}\right\vert ^{2}. \label{ppp1}%
\end{align}

Therefore, the membership matrix (\ref{ma}) can be defined as a linear
combination of the projection operators with the membership amplitudes as
coefficients%
\begin{equation}
\mathsf{M}\left(  \alpha_{0},\alpha_{1}\right)  =\alpha_{0}\mathsf{P}%
_{0}+\alpha_{1}\mathsf{P}_{1}.
\end{equation}

We compute%
\begin{equation}
\mathsf{M}\left(  \alpha_{0},\alpha_{1}\right)  \left\vert \psi_{ob}^{\left(
2\right)  }\right\rangle =\alpha_{0}^{2}a_{0}\left\vert 0\right\rangle
+\alpha_{1}^{2}a_{1}\left\vert 1\right\rangle .
\end{equation}

We can therefore treat the application of the membership matrix (\ref{ps}) as providing the
origin of a reversible but non-unitary \textquotedblleft obscure
measurement\textquotedblright\ on the standard qubit to obtain an obscure
qubit (cf. the \textquotedblleft mirror measurement\textquotedblright%
\ \cite{bat/ziz,ziz05} and also the origin of ordinary qubit states on the fuzzy
sphere \cite{ziz/pes}).

An obscure analog of the density operator (for a pure state) is the following form for the density
matrix in the vector representation%
\begin{equation}
\rho_{ob}^{\left(  2\right)  }=\left\vert \psi_{ob}^{\left(  2\right)
}\right\rangle \left\langle \psi_{ob}^{\left(  2\right)  }\right\vert =\left(
\begin{array}
[c]{cc}%
\alpha_{0}^{2}\left\vert a_{0}\right\vert ^{2} & \alpha_{0}a_{0}^{\ast}%
\alpha_{1}a_{1}\\
\alpha_{0}a_{0}\alpha_{1}a_{1}^{\ast} & \alpha_{1}^{2}\left\vert
a_{1}\right\vert ^{2}%
\end{array}
\right)  \label{r}%
\end{equation}
with the obvious standard singularity property $\det\rho_{ob}^{\left(
2\right)  }=0$. But $\operatorname*{tr}\rho_{ob}^{\left(  2\right)  }%
=\alpha_{0}^{2}\left\vert a_{0}\right\vert ^{2}+\alpha_{1}^{2}\left\vert
a_{1}\right\vert ^{2}\neq1$, and here there is no idempotence $\left(
\rho_{ob}^{\left(  2\right)  }\right)  ^{2}\neq\rho_{ob}^{\left(  2\right)  }%
$, which distincts $\rho_{ob}^{\left(  2\right)  }$ from the standard density operator.

\section{Kronecker obscure qubits}

We next introduce an analog of
quantum superposition for membership amplitudes, called \textquotedblleft obscure
superposition\textquotedblright\ (cf. \cite{cun/sha/tof/dub}, and also
\cite{tof/dub}).

Quantum amplitudes and membership amplitudes will here be considered separately in
order to define an \textquotedblleft obscure qubit\textquotedblright taking the
form of a \textquotedblleft double superposition\textquotedblright\ (cf.
(\ref{p2}), and a generalized  analog for qudits (\ref{pn}) is straightforward)%
\begin{equation}
\left\vert \mathbf{\Psi}_{ob}\right\rangle =\frac{\mathsf{A}_{0}\left\vert
\mathsf{0}\right\rangle +\mathsf{A}_{1}\left\vert \mathsf{1}\right\rangle
}{\sqrt{2}}, \label{pa}%
\end{equation}
where the two-dimensional \textquotedblleft vectors\textquotedblright%
\begin{equation}
\mathsf{A}_{0,1}=\left[
\begin{array}
[c]{c}%
a_{0,1}\\
\alpha_{0,1}%
\end{array}
\right]  \label{a01}%
\end{equation}
are the (double) \textquotedblleft obscure-quantum
amplitudes\textquotedblright\ of the generalized states $\left\vert
\mathsf{0}\right\rangle $, $\left\vert \mathsf{1}\right\rangle $. For the
conjugate of an obscure qubit we put (informally)%
\begin{equation}
\left\langle \mathbf{\Psi}_{ob}\right\vert =\frac{\mathsf{A}_{0}^{\star
}\left\langle \mathsf{0}\right\vert +\mathsf{A}_{1}^{\star}\left\langle
\mathsf{1}\right\vert }{\sqrt{2}}, \label{pa1}%
\end{equation}
where we denote $\mathsf{A}_{0,1}^{\star}=\left[
\begin{array}
[c]{cc}%
a_{0,1}^{\ast} & \alpha_{0,1}%
\end{array}
\right]  $, such that $\mathsf{A}_{0,1}^{\star}\mathsf{A}_{0,1}=\left\vert
a_{0,1}\right\vert ^{2}+\alpha_{0,1}^{2}$. The (double) obscure qubit is
\textquotedblleft normalized\textquotedblright\ in such a way that, if
the conditions (\ref{n1})--(\ref{n2}) hold, then
\begin{equation}
\left\langle \mathbf{\Psi}_{ob}\mid\mathbf{\Psi}_{ob}\right\rangle
=\frac{\left\vert a_{0}\right\vert ^{2}+\left\vert a_{1}\right\vert ^{2}}%
{2}+\frac{\alpha_{0}^{2}+\alpha_{1}^{2}}{2}=1. \label{ppn}%
\end{equation}

A measurement should be made separately and independently in the
\textquotedblleft probability space\textquotedblright\ and the
\textquotedblleft membership space\textquotedblright\, which can be represented by using an analog of
the Kronecker product. Indeed, in the vector
representation (\ref{vr}) for the quantum states and for the direct product
amplitudes (\ref{a01}) we should have%
\begin{equation}
\left\vert \mathbf{\Psi}_{ob}\right\rangle _{\left(  0\right)  }=\frac
{1}{\sqrt{2}}\mathsf{A}_{0}\otimes_{K}\left(
\begin{array}
[c]{c}%
1\\
0
\end{array}
\right)  +\mathsf{A}_{1}\otimes_{K}\left(
\begin{array}
[c]{c}%
0\\
1
\end{array}
\right)  , \label{ps2}%
\end{equation}
where the (left) Kronecker product is defined by (see (\ref{vr}))%
\begin{align}
&\left[
\begin{array}
[c]{c}%
a\\
\alpha
\end{array}
\right]  \otimes_{K}\left(
\begin{array}
[c]{c}%
c\\
d
\end{array}
\right)  =\left[
\begin{array}
[c]{c}%
a\left(
\begin{array}
[c]{c}%
c\\
d
\end{array}
\right) \\
\alpha\left(
\begin{array}
[c]{c}%
c\\
d
\end{array}
\right)
\end{array}
\right]  =\left[
\begin{array}
[c]{c}%
a\left(  c\mathsf{e}_{0}+d\mathsf{e}_{1}\right) \\
\alpha\left(  c\mathsf{e}_{0}+d\mathsf{e}_{1}\right)
\end{array}
\right]  ,\label{aa}\\
& \mathsf{e}_{0}=\left(
\begin{array}
[c]{c}%
1\\
0
\end{array}
\right)  ,\ \mathsf{e}_{1}=\left(
\begin{array}
[c]{c}%
0\\
1
\end{array}
\right)  ,\ \mathsf{e}_{0,1}\in\mathcal{H}_{q}^{\left(  2\right)  }.\nonumber
\end{align}

Informally, the wave function of the obscure qubit, in the vector
representation, now \textquotedblleft lives\textquotedblright\ in the
four-dimensional space of (\ref{aa}) which has two two-dimensional spaces as blocks.
The upper block, the quantum subspace, is the ordinary Hilbert space
$\mathcal{H}_{q}^{\left(  2\right)  }$, but the lower block should have special (fuzzy) properties, if it is
treated as an obscure (membership) subspace $\mathcal{V}_{memb}^{\left(
2\right)  }$. Thus, the four-dimensional space,
where \textquotedblleft lives\textquotedblright\ $\left\vert \Psi
_{ob}^{\left(  2\right)  }\right\rangle $, is not an ordinary tensor product of
vector spaces, because of (\ref{aa}), and the \textquotedblleft
vector\textquotedblright\ $\mathsf{A}$ on the l.h.s. has entries of different
natures, that is the quantum amplitudes $a_{0,1}$ and the membership
amplitudes $\alpha_{0,1}$. Despite the unit vectors in $\mathcal{H}%
_{q}^{\left(  2\right)  }$ and $\mathcal{V}_{memb}^{\left(  2\right)  }$
having the same form (\ref{vr}), they belong to different spaces (as they are vector
spaces over different fields). Therefore, instead of (\ref{aa}) we introduce a
\textquotedblleft Kronecker-like product\textquotedblright\ $\tilde{\otimes
}_{K}$ by%
\begin{align}
\left[
\begin{array}
[c]{c}%
a\\
\alpha
\end{array}
\right]  \tilde{\otimes}_{K}\left(
\begin{array}
[c]{c}%
c\\
d
\end{array}
\right)   &  =\left[
\begin{array}
[c]{c}%
a\left(  c\mathsf{e}_{0}+d\mathsf{e}_{1}\right) \\
\alpha\left(  c\varepsilon_{0}+d\varepsilon_{1}\right)
\end{array}
\right]  ,\label{akc}\\
\mathsf{e}_{0}  &  =\left(
\begin{array}
[c]{c}%
1\\
0
\end{array}
\right)  ,\ \ \ \mathsf{e}_{1}=\left(
\begin{array}
[c]{c}%
0\\
1
\end{array}
\right)  ,\ \ \ \ \mathsf{e}_{0,1}\in\mathcal{H}_{q}^{\left(  2\right)
},\label{e1}\\
\varepsilon_{0}  &  =\left(
\begin{array}
[c]{c}%
1\\
0
\end{array}
\right)  ^{\left(  \mu\right)  },\ \ \ \varepsilon_{1}=\left(
\begin{array}
[c]{c}%
0\\
1
\end{array}
\right)  ^{\left(  \mu\right)  },\ \ \ \ \varepsilon_{0,1}\in\mathcal{V}%
_{memb}^{\left(  2\right)  }. \label{e2}%
\end{align}

In this way, the obscure qubit (\ref{pa}) can be presented in the from%
\begin{align}
\left\vert \mathbf{\Psi}_{ob}\right\rangle =\frac{1}{\sqrt{2}}\left[
\begin{array}
[c]{c}%
a_{0}\left(
\begin{array}
[c]{c}%
1\\
0
\end{array}
\right) \\
\alpha_{0}\left(
\begin{array}
[c]{c}%
1\\
0
\end{array}
\right)  ^{\left(  \mu\right)  }%
\end{array}
\right]  +\frac{1}{\sqrt{2}}\left[
\begin{array}
[c]{c}%
a_{1}\left(
\begin{array}
[c]{c}%
0\\
1
\end{array}
\right) \\
\alpha_{1}\left(
\begin{array}
[c]{c}%
0\\
1
\end{array}
\right)  ^{\left(  \mu\right)  }%
\end{array}
\right]\nonumber\\
=\frac{1}{\sqrt{2}}\left[
\begin{array}
[c]{c}%
a_{0}\mathsf{e}_{0}\\
\alpha_{0}\varepsilon_{0}%
\end{array}
\right]  +\frac{1}{\sqrt{2}}\left[
\begin{array}
[c]{c}%
a_{1}\mathsf{e}_{1}\\
\alpha_{1}\varepsilon_{1}%
\end{array}
\right]  . \label{ps3}%
\end{align}

Therefore, we call the double obscure qubit (\ref{ps3}) a \textquotedblleft
Kronecker obscure qubit\textquotedblright\ to distinguish it from the obscure
qubit (\ref{p2}). It can be also presented using the Hadamard product
(the element-wise or Schur product)%
\begin{equation}
\left[
\begin{array}
[c]{c}%
a\\
\alpha
\end{array}
\right]  \otimes_{H}\left(
\begin{array}
[c]{c}%
c\\
d
\end{array}
\right)  =\left[
\begin{array}
[c]{c}%
ac\\
\alpha d
\end{array}
\right]  \label{h}%
\end{equation}
in the following form%
\begin{equation}
\left\vert \mathbf{\Psi}_{ob}\right\rangle =\frac{1}{\sqrt{2}}\mathsf{A}%
_{0}\otimes_{H}\mathsf{E}_{0}+\frac{1}{\sqrt{2}}\mathsf{A}_{1}\otimes
_{H}\mathsf{E}_{1}, \label{ph}%
\end{equation}
where the unit vectors of the total four-dimensional space are%
\begin{equation}
\mathsf{E}_{0,1}=\left[
\begin{array}
[c]{c}%
\mathsf{e}_{0,1}\\
\varepsilon_{0,1}%
\end{array}
\right]  \in\mathcal{H}_{q}^{\left(  2\right)  }\times\mathcal{V}%
_{memb}^{\left(  2\right)  }. \label{ee}%
\end{equation}

The probabilities $p_{0,1}$ and membership functions $\mu_{0,1}$ of the states
$\left\vert \mathsf{0}\right\rangle $ and $\left\vert \mathsf{1}\right\rangle
$ are computed through the corresponding amplitudes by (\ref{fb}) and
(\ref{mm})%
\begin{equation}
p_{i}=\left\vert a_{i}\right\vert ^{2},\ \ \ \ \mu_{i}=F_{\mu_{i}}\left(
\alpha_{0},\alpha_{1}\right)  ,\ \ \ i=0,1, \label{pm}%
\end{equation}
and in the particular case by (\ref{mm1}) satisfying (\ref{n2}).

By way of example, consider a Kronecker obscure qubit (with a real quantum
part) with  probability $p$ and membership function $\mu$ (measure of
\textquotedblleft trust\textquotedblright) of the state $\left\vert
\mathsf{0}\right\rangle $, and of the state $\left\vert
\mathsf{1}\right\rangle $ given by $1-p$ and $1-\mu$ respectively. In the
model (\ref{f1})--(\ref{f2}) for $\mu_{i}$ (which is not Born-like) we obtain%
\begin{align}
\left\vert \mathbf{\Psi}_{ob}\right\rangle &=\frac{1}{\sqrt{2}}\left[
\begin{array}
[c]{c}%
\left(
\begin{array}
[c]{c}%
\sqrt{p}\\
0
\end{array}
\right) \\
\left(
\begin{array}
[c]{c}%
\cos\dfrac{\pi}{2}\mu\\
0
\end{array}
\right)  ^{\left(  \mu\right)  }%
\end{array}
\right]  +\frac{1}{\sqrt{2}}\left[
\begin{array}
[c]{c}%
\left(
\begin{array}
[c]{c}%
0\\
\sqrt{1-p}%
\end{array}
\right) \\
\left(
\begin{array}
[c]{c}%
0\\
\sin\dfrac{\pi}{2}\mu
\end{array}
\right)  ^{\left(  \mu\right)  }%
\end{array}
\right] \nonumber \\
&=\frac{1}{\sqrt{2}}\left[
\begin{array}
[c]{c}%
\mathsf{e}_{0}\sqrt{p}\\
\varepsilon_{0}\cos\dfrac{\pi}{2}\mu
\end{array}
\right]  +\frac{1}{\sqrt{2}}\left[
\begin{array}
[c]{c}%
\mathsf{e}_{1}\sqrt{1-p}\\
\varepsilon_{1}\sin\dfrac{\pi}{2}\mu
\end{array}
\right]  , \label{ppm}%
\end{align}
where $\mathsf{e}_{i}$ and $\varepsilon_{i}$ are unit vectors defined in
(\ref{e1}) and (\ref{e2}).

This can be compared e.g. with the \textquotedblleft
classical-quantum\textquotedblright\ approach (\ref{s}) and
\cite{man06,mar/vis/rei13}, in which the elements of columns are multiplied,
while we consider them independently and separately.

\section{Obscure-quantum measurement}

Let us consider the case of one Kronecker obscure qubit register $L=1$ (see
(\ref{ps2})), or using (\ref{aa}) in the vector representation (\ref{ps3}).
The standard (double)\ orthogonal commuting projection operators,
\textquotedblleft Kronecker projections\textquotedblright\ are (cf.
(\ref{pp}))%
\begin{equation}
\mathbf{P}_{0}=\left[
\begin{array}
[c]{cc}%
\mathsf{P}_{0} & \mathsf{0}\\
\mathsf{0} & \mathsf{P}_{0}^{\left(  \mu\right)  }%
\end{array}
\right]  ,\ \ \ \ \mathbf{P}_{1}=\left[
\begin{array}
[c]{cc}%
\mathsf{P}_{1} & \mathsf{0}\\
\mathsf{0} & \mathsf{P}_{1}^{\left(  \mu\right)  }%
\end{array}
\right]  , \label{pp01}%
\end{equation}
where $\mathsf{0}$ is the $2\times2$ zero matrix, and $\mathsf{P}%
_{0,1}^{\left(  \mu\right)  }$ are the projections in the membership subspace
$\mathcal{V}_{memb}^{\left(  2\right)  }$ (of the same form as the ordinary
quantum projections $\mathsf{P}_{0,1}$ (\ref{pp}))%
\begin{align}
\mathsf{P}_{0}^{\left(  \mu\right)  }  &  =\left(
\begin{array}
[c]{cc}%
1 & 0\\
0 & 0
\end{array}
\right)  ^{\left(  \mu\right)  },\ \ \ \ \mathsf{P}_{1}^{\left(  \mu\right)
}=\left(
\begin{array}
[c]{cc}%
0 & 0\\
0 & 1
\end{array}
\right)  ^{\left(  \mu\right)  },\ \ \ \mathsf{P}_{0}^{\left(  \mu\right)
},\mathsf{P}_{1}^{\left(  \mu\right)  }\in\operatorname*{End}\mathcal{V}%
_{memb}^{\left(  2\right)  },\\
\mathsf{P}_{0}^{\left(  \mu\right)  2}  &  =\mathsf{P}_{0}^{\left(
\mu\right)  },\ \ \mathsf{P}_{1}^{\left(  \mu\right)  2}=\mathsf{P}%
_{1}^{\left(  \mu\right)  },\ \ \mathsf{P}_{0}^{\left(  \mu\right)
}\mathsf{P}_{1}^{\left(  \mu\right)  }=\mathsf{P}_{1}^{\left(  \mu\right)
}\mathsf{P}_{0}^{\left(  \mu\right)  }=\mathsf{0}.
\end{align}

For the double\ projections we have (cf. (\ref{pp1}))%
\begin{equation}
\mathbf{P}_{0}^{2}=\mathbf{P}_{0},\ \ \mathbf{P}_{1}^{2}=\mathbf{P}%
_{1},\ \ \mathbf{P}_{0}\mathbf{P}_{1}=\mathbf{P}_{1}\mathbf{P}_{0}=\mathbf{0},
\label{p2p}%
\end{equation}
where $\mathbf{0}$ is the $4\times4$ zero matrix, and $\mathbf{P}_{0,1}$ act
on the Kronecker qubit (\ref{pp01}) in the standard way (cf. (\ref{ppp}))%
\begin{align}
\mathbf{P}_{0}\left\vert \mathbf{\Psi}_{ob}\right\rangle  &  =\frac{1}%
{\sqrt{2}}\left[
\begin{array}
[c]{c}%
a_{0}\left(
\begin{array}
[c]{c}%
1\\
0
\end{array}
\right) \\
\alpha_{0}\left(
\begin{array}
[c]{c}%
1\\
0
\end{array}
\right)  ^{\left(  \mu\right)  }%
\end{array}
\right]  =\frac{1}{\sqrt{2}}\left[
\begin{array}
[c]{c}%
a_{0}\mathsf{e}_{0}\\
\alpha_{0}\varepsilon_{0}%
\end{array}
\right]  =\frac{1}{\sqrt{2}}\mathsf{A}_{0}\otimes_{H}\mathsf{E}_{0}%
,\label{pps}\\
\mathbf{P}_{1}\left\vert \mathbf{\Psi}_{ob}\right\rangle  &  =\frac{1}%
{\sqrt{2}}\left[
\begin{array}
[c]{c}%
a_{1}\left(
\begin{array}
[c]{c}%
0\\
1
\end{array}
\right) \\
\alpha_{1}\left(
\begin{array}
[c]{c}%
0\\
1
\end{array}
\right)  ^{\left(  \mu\right)  }%
\end{array}
\right]  =\frac{1}{\sqrt{2}}\left[
\begin{array}
[c]{c}%
a_{1}\mathsf{e}_{1}\\
\alpha_{1}\varepsilon_{1}%
\end{array}
\right]  =\frac{1}{\sqrt{2}}\mathsf{A}_{1}\otimes_{H}\mathsf{E}_{1}.
\end{align}

Observe that for Kronecker qubits there exist in addition to (\ref{pp01}) the
following orthogonal commuting projection operators%
\begin{equation}
\mathbf{P}_{01}=\left[
\begin{array}
[c]{cc}%
\mathsf{P}_{0} & \mathsf{0}\\
\mathsf{0} & \mathsf{P}_{1}^{\left(  \mu\right)  }%
\end{array}
\right]  ,\ \ \ \ \mathbf{P}_{10}=\left[
\begin{array}
[c]{cc}%
\mathsf{P}_{1} & \mathsf{0}\\
\mathsf{0} & \mathsf{P}_{0}^{\left(  \mu\right)  }%
\end{array}
\right]  , \label{ppcr}%
\end{equation}
and we call these the \textquotedblleft crossed\textquotedblright\ double
projections. They satisfy the same relations as (\ref{p2p})%
\begin{equation}
\mathbf{P}_{01}^{2}=\mathbf{P}_{01},\ \ \mathbf{P}_{10}^{2}=\mathbf{P}%
_{10},\ \ \mathbf{P}_{01}\mathbf{P}_{10}=\mathbf{P}_{10}\mathbf{P}%
_{01}=\mathbf{0},
\end{equation}
but act on the obscure qubit in a different (\textquotedblleft
mixing\textquotedblright) way than (\ref{pps}) i.e.%
\begin{align}
&\mathbf{P}_{01}\left\vert \mathbf{\Psi}_{ob}\right\rangle =\frac{1}{\sqrt{2}%
}\left[
\begin{array}
[c]{c}%
a_{0}\left(
\begin{array}
[c]{c}%
1\\
0
\end{array}
\right) \\
\alpha_{1}\left(
\begin{array}
[c]{c}%
0\\
1
\end{array}
\right)
\end{array}
\right]  =\frac{1}{\sqrt{2}}\left[
\begin{array}
[c]{c}%
a_{0}\mathsf{e}_{0}\\
\alpha_{1}\varepsilon_{1}%
\end{array}
\right]  ,\\
&\mathbf{P}_{10}\left\vert \mathbf{\Psi}_{ob}\right\rangle
=\frac{1}{\sqrt{2}}\left[
\begin{array}
[c]{c}%
a_{1}\left(
\begin{array}
[c]{c}%
0\\
1
\end{array}
\right) \\
\alpha_{0}\left(
\begin{array}
[c]{c}%
1\\
0
\end{array}
\right)
\end{array}
\right]  =\frac{1}{\sqrt{2}}\left[
\begin{array}
[c]{c}%
a_{1}\mathsf{e}_{1}\\
\alpha_{0}\varepsilon_{0}%
\end{array}
\right]  . \label{pps1}%
\end{align}

The multiplication of the crossed double projections (\ref{ppcr}) and the
double projections (\ref{pp01}) is given by%
\begin{align}
\mathbf{P}_{01}\mathbf{P}_{0}  &  =\mathbf{P}_{0}\mathbf{P}_{01}=\left[
\begin{array}
[c]{cc}%
\mathsf{P}_{0} & \mathsf{0}\\
\mathsf{0} & \mathsf{0}%
\end{array}
\right]  \equiv\mathbf{Q}_{0},\ \ \ \ \mathbf{P}_{01}\mathbf{P}_{1}%
=\mathbf{P}_{1}\mathbf{P}_{01}=\left[
\begin{array}
[c]{cc}%
\mathsf{0} & \mathsf{0}\\
\mathsf{0} & \mathsf{P}_{1}^{\left(  \mu\right)  }%
\end{array}
\right]  \equiv\mathbf{Q}_{1}^{\left(  \mu\right)  },\label{pq1}\\
\mathbf{P}_{10}\mathbf{P}_{0}  &  =\mathbf{P}_{0}\mathbf{P}_{10}=\left[
\begin{array}
[c]{cc}%
\mathsf{0} & \mathsf{0}\\
\mathsf{0} & \mathsf{P}_{0}^{\left(  \mu\right)  }%
\end{array}
\right]  \equiv\mathbf{Q}_{0}^{\left(  \mu\right)  },\ \ \ \ \mathbf{P}%
_{10}\mathbf{P}_{1}=\mathbf{P}_{1}\mathbf{P}_{10}=\left[
\begin{array}
[c]{cc}%
\mathsf{P}_{1} & \mathsf{0}\\
\mathsf{0} & \mathsf{0}%
\end{array}
\right]  \equiv\mathbf{Q}_{1}, \label{pq2}%
\end{align}
where the operators $\mathbf{Q}_{0},\mathbf{Q}_{1}$ and $\mathbf{Q}%
_{0}^{\left(  \mu\right)  },\mathbf{Q}_{1}^{\left(  \mu\right)  }$ satisfy%
\begin{align}
\mathbf{Q}_{0}^{2}  &  =\mathbf{Q}_{0},\ \ \mathbf{Q}_{1}^{2}=\mathbf{Q}%
_{1},\ \ \mathbf{Q}_{1}\mathbf{Q}_{0}=\mathbf{Q}_{0}\mathbf{Q}_{1}%
=\mathbf{0},\\
\mathbf{Q}_{0}^{\left(  \mu\right)  2}  &  =\mathbf{Q}_{0}^{\left(
\mu\right)  },\ \ \mathbf{Q}_{1}^{\left(  \mu\right)  2}=\mathbf{Q}%
_{1}^{\left(  \mu\right)  },\ \ \mathbf{Q}_{1}^{\left(  \mu\right)
}\mathbf{Q}_{0}^{\left(  \mu\right)  }=\mathbf{Q}_{0}^{\left(  \mu\right)
}\mathbf{Q}_{1}^{\left(  \mu\right)  }=\mathbf{0},\\
\mathbf{Q}_{1}^{\left(  \mu\right)  }\mathbf{Q}_{0}  &  =\mathbf{Q}%
_{0}^{\left(  \mu\right)  }\mathbf{Q}_{1}=\mathbf{Q}_{1}\mathbf{Q}%
_{0}^{\left(  \mu\right)  }=\mathbf{Q}_{0}\mathbf{Q}_{1}^{\left(  \mu\right)
}=\mathbf{0},
\end{align}
and we call these \textquotedblleft half Kronecker (double)
projections\textquotedblright.

The relations above imply that the process of measurement when using Kronecker
obscure qubits (i.e. for quantum computation with truth or membership) is more
complicated than in the standard case.

To show this, let us calculate the \textquotedblleft obscure\textquotedblright%
\ analogs of expected values for the projections above. Using the notation%
\begin{equation}
\mathbf{\bar{A}}\equiv\left\langle \mathbf{\Psi}_{ob}\right\vert
\mathbf{A}\left\vert \mathbf{\Psi}_{ob}\right\rangle .
\end{equation}

Then, using (\ref{pa})--(\ref{pa1}) for the projection operators
$\mathbf{P}_{i}$, $\mathbf{P}_{ij}$, $\mathbf{Q}_{i}$, $\mathbf{Q}%
_{i}^{\left(  \mu\right)  }$, $i,j=0,1$, $i\neq j$, we obtain (cf.
(\ref{ppp1}))%
\begin{align}
\mathbf{\bar{P}}_{i}  &  =\frac{\left\vert a_{i}\right\vert ^{2}+\alpha
_{i}^{2}}{2},\ \ \ \ \mathbf{\bar{P}}_{ij}=\frac{\left\vert a_{i}\right\vert
^{2}+\alpha_{j}^{2}}{2},\\
\mathbf{\bar{Q}}_{i}  &  =\frac{\left\vert a_{i}\right\vert ^{2}}%
{2},\ \ \ \ \ \ \ \ \ \ \mathbf{\bar{Q}}_{i}^{\left(  \mu\right)  }%
=\frac{\alpha_{i}^{2}}{2}.
\end{align}

So follows the relation between the \textquotedblleft
obscure\textquotedblright\ analogs of expected values of the projections%
\begin{equation}
\mathbf{\bar{P}}_{i}=\mathbf{\bar{Q}}_{i}+\mathbf{\bar{Q}}_{i}^{\left(
\mu\right)  },\ \ \ \ \mathbf{\bar{P}}_{ij}=\mathbf{\bar{Q}}_{i}%
+\mathbf{\bar{Q}}_{j}^{\left(  \mu\right)  }.
\end{equation}

Taking the \textquotedblleft ket\textquotedblright\ corresponding to the \textquotedblleft
bra\textquotedblright\ Kronecker qubit (\ref{ps3}) in the form%
\begin{equation}
\left\langle \mathbf{\Psi}_{ob}\right\vert =\frac{1}{\sqrt{2}}\left[
\begin{array}
[c]{cc}%
a_{0}^{\ast}\left(
\begin{array}
[c]{cc}%
1 & 0
\end{array}
\right)  , & \alpha_{0}\left(
\begin{array}
[c]{cc}%
1 & 0
\end{array}
\right)
\end{array}
\right]  +\frac{1}{\sqrt{2}}\left[
\begin{array}
[c]{cc}%
a_{1}^{\ast}\left(
\begin{array}
[c]{cc}%
0 & 1
\end{array}
\right)  , & \alpha_{1}\left(
\begin{array}
[c]{cc}%
0 & 1
\end{array}
\right)
\end{array}
\right]  ,
\end{equation}
a Kronecker\ ($4\times4$) obscure analog of the density matrix for a pure
state is given by (cf. (\ref{r}))%
\begin{equation}
\mathbf{\rho}_{ob}^{\left(  2\right)  }=\left\vert \mathbf{\Psi}%
_{ob}\right\rangle \left\langle \mathbf{\Psi}_{ob}\right\vert =\frac{1}%
{2}\left(
\begin{array}
[c]{cccc}%
\left\vert a_{0}\right\vert ^{2} & a_{0}a_{1}^{\ast} & a_{0}\alpha_{0} &
a_{0}\alpha_{1}\\
a_{1}a_{0}^{\ast} & \left\vert a_{1}\right\vert ^{2} & a_{1}\alpha_{0} &
a_{1}\alpha_{1}\\
\alpha_{0}a_{0}^{\ast} & \alpha_{0}a_{1}^{\ast} & \alpha_{0}^{2} & \alpha
_{0}\alpha_{1}\\
\alpha_{1}a_{0}^{\ast} & \alpha_{1}a_{1}^{\ast} & \alpha_{0}\alpha_{1} &
\alpha_{1}^{2}%
\end{array}
\right)  .\label{rr}%
\end{equation}

If the Born rule for the membership functions (\ref{mm1}) and the conditions
(\ref{n1})--(\ref{n2}) are satisfied, the density matrix (\ref{rr}) is
non-invertible, because $\det\mathbf{\rho}_{ob}^{\left(  2\right)  }=0$ and
has unit trace $\operatorname*{tr}\mathbf{\rho}_{ob}^{\left(  2\right)  }=1$,
but is not idempotent $\left(  \mathbf{\rho}_{ob}^{\left(  2\right)  }\right)
^{2}\neq\mathbf{\rho}_{ob}^{\left(  2\right)  }$ (as it holds for the ordinary
quantum density matrix \cite{nie/chu}).

\section{Kronecker obscure-quantum gates}

In general, (double)\ \textquotedblleft obscure-quantum
computation\textquotedblright\ with $L$ Kronecker obscure qubits (or qudits)
can be performed by a product of unitary (block) matrices $\mathbf{U}$ of the
(double size to the standard one) size $2\times\left(  2^{L}\times
2^{L}\right)  $ (or $2\times\left(  n^{L}\times n^{L}\right)  $),
$\mathbf{U}^{\dag}\mathbf{U}=\mathbf{I}$ (here $\mathbf{I}$ is the unit matrix
of the same size as $\mathbf{U}$). We can also call such computation a
\textquotedblleft quantum computation with truth\textquotedblright\ (or with membership).

Let us consider obscure-quantum computation with one Kronecker obscure qubit.
Informally, we can present the Kronecker obscure qubit (\ref{ps3}) in the form%
\begin{equation}
\left\vert \mathbf{\Psi}_{ob}\right\rangle =\left[
\begin{array}
[c]{c}%
\frac{1}{\sqrt{2}}\left(
\begin{array}
[c]{c}%
a_{0}\\
a_{1}%
\end{array}
\right) \\
\frac{1}{\sqrt{2}}\left(
\begin{array}
[c]{c}%
\alpha_{0}\\
\alpha_{1}%
\end{array}
\right)  ^{\left(  \mu\right)  }%
\end{array}
\right]  . \label{p2a}%
\end{equation}

Thus, the state $\left\vert \mathbf{\Psi}_{ob}\right\rangle $ can be
interpreted as a \textquotedblleft vector\textquotedblright\ in the direct
product (not tensor product) space $\mathcal{H}_{q}^{\left(  2\right)  }%
\times\mathcal{V}_{memb}^{\left(  2\right)  }$, where $\mathcal{H}%
_{q}^{\left(  2\right)  }$ is the standard two-dimenional Hilbert space of the
qubit, and $\mathcal{V}_{memb}^{\left(  2\right)  }$ can be treated as the
\textquotedblleft membership space\textquotedblright\ which has a different
nature from the qubit space and can have a more complex structure. For
discussion of such spaces, see, e.g. \cite{dub/ngu/pra,belohl,smith,zim11}. In
general, one can consider obscure-quantum computation as a set of abstract
computational rules, independently of the introduction of the corresponding spaces.

An obscure-quantum gate will be defined as an elementary transformation on an obscure qubit (\ref{p2a}) and is performed by unitary (block) matrices of
size $4\times4$ (over $\mathbb{C}$) acting in the total space $\mathcal{H}%
_{q}^{\left(  2\right)  }\times\mathcal{V}_{memb}^{\left(  2\right)  }$
\begin{align}
\mathbf{U} &  =\left(
\begin{array}
[c]{cc}%
\mathsf{U} & \mathsf{0}\\
\mathsf{0} & \mathsf{U}^{\left(  \mu\right)  }%
\end{array}
\right)  ,\ \ \ \mathbf{UU}^{\dag}=\mathbf{U}^{\dag}\mathbf{U}=\mathbf{I}%
,\label{u}\\
\mathsf{UU}^{\dag} &  =\mathsf{U}^{\dag}\mathsf{U}=\mathsf{I},\ \ \mathsf{U}%
^{\left(  \mu\right)  }\mathsf{U}^{\left(  \mu\right)  \dag}=\mathsf{U}%
^{\left(  \mu\right)  \dag}\mathsf{U}^{\left(  \mu\right)  }=\mathsf{I}%
,\ \ \ \mathsf{U}\in\operatorname*{End}\mathcal{H}_{q}^{\left(  2\right)
},\mathsf{U}^{\left(  \mu\right)  }\in\operatorname*{End}\mathcal{V}%
_{memb}^{\left(  2\right)  },
\end{align}
where $\mathbf{I}$ is the unit $4\times4$ matrix, $\mathsf{I}$ is the unit
$2\times2$ matrix, $\mathsf{U}$ and $\mathsf{U}^{\left(  \mu\right)  }$ are
unitary $2\times2$ matrices acting on the probability and membership
\textquotedblleft subspaces\textquotedblright\ respectively. The matrix
$\mathsf{U}$ (over $\mathbb{C}$) will be called a quantum gate, and we call the
matrix $\mathsf{U}^{\left(  \mu\right)  }$ (over $\mathbb{R}$) an
\textquotedblleft obscure gate\textquotedblright. We assume that the obscure
gates $\mathsf{U}^{\left(  \mu\right)  }$ are of the same shape as the
standard quantum gates, but they act in the other (membership) space and have only real elements (see, e.g. \cite{nie/chu}). In
this case, an obscure-quantum gate is characterized by the pair $\left\{
\mathsf{U},\mathsf{U}^{\left(  \mu\right)  }\right\}  $, where the components
are known gates (in various combinations), e.g., for one qubit gates:
\textsf{Hadamard}, \textsf{Pauli-X }($\mathtt{NOT}$\texttt{)},\textsf{Y}%
,\textsf{Z }(or two qubit gates e.g. $\mathtt{CNOT}$, $\mathtt{SWAP}$, etc.).
The transformed qubit then becomes (informally)%
\begin{equation}
\mathbf{U}\left\vert \mathbf{\Psi}_{ob}\right\rangle =\left[
\begin{array}
[c]{c}%
\frac{1}{\sqrt{2}}\mathsf{U}\left(
\begin{array}
[c]{c}%
a_{0}\\
a_{1}%
\end{array}
\right)  \\
\frac{1}{\sqrt{2}}\mathsf{U}^{\left(  \mu\right)  }\left(
\begin{array}
[c]{c}%
\alpha_{0}\\
\alpha_{1}%
\end{array}
\right)  ^{\left(  \mu\right)  }%
\end{array}
\right]  .\label{up}%
\end{equation}
Thus the quantum and the membership parts
are transformed independently for the block diagonal form (\ref{u}). Some examples of this can be found, e.g.,
in \cite{dom/fre,man06,mar/vis/rei13}. Differences between the parts
were mentioned in \cite{kre/koh/kim}. In this case, an obscure-quantum network
is \textquotedblleft physically\textquotedblright\ realised by a device performing
elementary operations in sequence on obscure qubits (by a product of matrices), such
that the quantum and membership parts are synchronized in time (for a discussion of the
obscure part of such physical devices, see
\cite{hir/oza89,koc/hir,virant,kosko}). Then, the result of the
obscure-quantum computation consists of the quantum probabilities
of the states together with the calculated \textquotedblleft level of truth\textquotedblright\ for
each of them (see, e.g. \cite{bol18}).

For example, the obscure-quantum gate $\mathbf{U}_{\mathsf{H,}\mathtt{NOT}%
}=\left\{  \text{\textsf{Hadamard}},\mathtt{NOT}\right\}  $ acts on the state
$\mathsf{E}_{0}$ (\ref{ee}) as follows%
\begin{equation}
\mathbf{U}_{\mathsf{H,}\mathtt{NOT}}\mathsf{E}_{0}=\mathbf{U}_{\mathsf{H,}%
\mathtt{NOT}}\left[
\begin{array}
[c]{c}%
\left(
\begin{array}
[c]{c}%
1\\
0
\end{array}
\right) \\
\left(
\begin{array}
[c]{c}%
1\\
0
\end{array}
\right)  ^{\left(  \mu\right)  }%
\end{array}
\right]  =\left[
\begin{array}
[c]{c}%
\frac{1}{\sqrt{2}}\left(
\begin{array}
[c]{c}%
1\\
1
\end{array}
\right) \\
\left(
\begin{array}
[c]{c}%
0\\
1
\end{array}
\right)  ^{\left(  \mu\right)  }%
\end{array}
\right]  =\left[
\begin{array}
[c]{c}%
\frac{1}{\sqrt{2}}\left(  \mathsf{e}_{0}+\mathsf{e}_{1}\right) \\
\varepsilon_{1}%
\end{array}
\right]  .
\end{equation}

It would be interesting to consider the case when $\mathbf{U}$ (\ref{u}) is
not block diagonal and try to find possible \textquotedblleft
physical\textquotedblright\ interpretations of the non-diagonal blocks.

\section{Double entanglement}

Let us introduce a register consisting of two obscure qubits ($L=2$)
in the computational basis $\left\vert \mathsf{ij}^{\prime}\right\rangle
=\left\vert \mathsf{i}\right\rangle \otimes\left\vert \mathsf{j}^{\prime
}\right\rangle $ as follows
\begin{equation}
\left\vert \mathbf{\Psi}_{ob}^{\left(  n=2\right)  }\left(  L=2\right)
\right\rangle =\left\vert \mathbf{\Psi}_{ob}\left(  2\right)  \right\rangle
=\frac{\mathsf{B}_{00^{\prime}}\left\vert \mathsf{00}^{\prime}\right\rangle
+\mathsf{B}_{10^{\prime}}\left\vert \mathsf{10}^{\prime}\right\rangle
+\mathsf{B}_{01^{\prime}}\left\vert \mathsf{01}^{\prime}\right\rangle
+\mathsf{B}_{11^{\prime}}\left\vert \mathsf{11}^{\prime}\right\rangle }%
{\sqrt{2}}, \label{l2}%
\end{equation}
determined by two-dimensional \textquotedblleft
vectors\textquotedblright\ (encoding obscure-quantum amplitudes)%
\begin{equation}
\mathsf{B}_{ij^{\prime}}=\left[
\begin{array}
[c]{c}%
b_{ij^{\prime}}\\
\beta_{ij^{\prime}}%
\end{array}
\right]  ,\ \ \ \ i,j=0,1,\ \ \ j^{\prime}=0^{\prime},1^{\prime}, \label{bb}%
\end{equation}
where $b_{ij^{\prime}}\in\mathbb{C}$ are probability amplitudes for a set of pure states and
$\beta_{ij^{\prime}}\in\mathbb{R}$ are the corresponding membership amplitudes .
By analogy with (\ref{pa}) and (\ref{ppn}) the normalization factor in (\ref{l2}) is chosen so that%
\begin{equation}
\left\langle \mathbf{\Psi}_{ob}\left(  2\right)  \mid\mathbf{\Psi}_{ob}\left(
2\right)  \right\rangle =1,
\end{equation}
if (cf. (\ref{n1})--(\ref{n2}))%
\begin{align}
\left\vert b_{00^{\prime}}\right\vert ^{2}+\left\vert b_{10^{\prime}%
}\right\vert ^{2}+\left\vert b_{01^{\prime}}\right\vert ^{2}+\left\vert
b_{11^{\prime}}\right\vert ^{2}  &  =1,\label{bb1}\\
\beta_{00^{\prime}}^{2}+\beta_{10^{\prime}}^{2}+\beta_{01^{\prime}}^{2}%
+\beta_{11^{\prime}}^{2}  &  =1. \label{bb2}%
\end{align}

A state of two qubits is \textquotedblleft entangled\textquotedblright, if it
cannot be decomposed as a product of two one-qubit states, and otherwise
it is \textquotedblleft separable\textquotedblright\ (see, e.g.
\cite{nie/chu}). We define a product of two obscure qubits (\ref{pa}) as%
\begin{equation}
\left\vert \mathbf{\Psi}_{ob}\right\rangle \otimes\left\vert \mathbf{\Psi
}_{ob}^{\prime}\right\rangle =\frac{\mathsf{A}_{0}\otimes_{H}\mathsf{A}%
_{0}^{\prime}\left\vert \mathsf{00}^{\prime}\right\rangle +\mathsf{A}%
_{1}\otimes_{H}\mathsf{A}_{0}^{\prime}\left\vert \mathsf{10}^{\prime
}\right\rangle +\mathsf{A}_{0}\otimes_{H}\mathsf{A}_{1}^{\prime}\left\vert
\mathsf{01}^{\prime}\right\rangle +\mathsf{A}_{1}\otimes_{H}\mathsf{A}%
_{1}^{\prime}\left\vert \mathsf{11}^{\prime}\right\rangle }{2}, \label{p22}%
\end{equation}
where $\otimes_{H}$ is the Hadamard product (\ref{h}). Comparing (\ref{l2})
and (\ref{p22}) we obtain two sets of relations, for probability amplitudes
and for membership amplitudes%
\begin{align}
b_{ij^{\prime}}  &  =\frac{1}{\sqrt{2}}a_{i}a_{j^{\prime}},\label{ba}\\
\beta_{ij^{\prime}}  &  =\frac{1}{\sqrt{2}}\alpha_{i}\alpha_{j^{\prime}%
},\ \ \ \ i,j=0,1,\ \ \ j^{\prime}=0^{\prime},1^{\prime}. \label{ba1}%
\end{align}

In this case, the relations (\ref{n1})--(\ref{n2}) give (\ref{bb1}%
)--(\ref{bb2}).

Two obscure-quantum qubits are entangled, if their joint state (\ref{l2})
cannot be presented as a product of one qubit states (\ref{p22}), and in the
opposite case the states are called totally separable. It follows from
(\ref{ba})-(\ref{ba1}), that there are two general conditions for obscure
qubits to be entangled%
\begin{align}
b_{00^{\prime}}b_{11^{\prime}}  &  \neq b_{10^{\prime}}b_{01^{\prime}%
},\ \ \ \ \ \text{or }\det\mathbf{b}\neq0,\ \mathbf{b}=\left(
\begin{array}
[c]{cc}%
b_{00^{\prime}} & b_{01^{\prime}}\\
b_{10^{\prime}} & b_{11^{\prime}}%
\end{array}
\right)  ,\label{b1}\\
\beta_{00^{\prime}}\beta_{11^{\prime}}  &  \neq\beta_{10^{\prime}}%
\beta_{01^{\prime}},\ \ \ \ \text{or }\det\mathbf{\beta}\neq0,\ \mathbf{\beta
}=\left(
\begin{array}
[c]{cc}%
\beta_{00^{\prime}} & \beta_{01^{\prime}}\\
\beta_{10^{\prime}} & \beta_{11^{\prime}}%
\end{array}
\right)  \cdot\label{b2}%
\end{align}
The first equation (\ref{b1}) is the entanglement relation for the standard
qubit, while the second condition (\ref{b2}) is for the membership amplitudes
of the two obscure qubit joint state (\ref{l2}). The presence of two different conditions
(\ref{b1})--(\ref{b2}) leads to new additional possibilities (which do not exist for
ordinary qubits) for \textquotedblleft partial\textquotedblright%
\ entanglement (or \textquotedblleft partial\textquotedblright\ separability),
when only one of them is fulfilled. In this case, the states can be
entangled in one subspace (quantum or membership) but not in the other.

The measure of entanglement is numerically characterized by the concurrence.
Taking into account the two conditions (\ref{b1})--(\ref{b2}), we propose to
generalize the notion of concurrence for two obscure qubits in two ways.
First, we introduce the \textquotedblleft vector obscure
concurrence\textquotedblright%
\begin{equation}
\mathsf{C}_{vect}=\left[
\begin{array}
[c]{c}%
C_{q}\\
C^{\left(  \mu\right)  }%
\end{array}
\right]  =2\left[
\begin{array}
[c]{c}%
\left\vert \det\mathbf{b}\right\vert \\
\left\vert \det\beta\right\vert
\end{array}
\right]  ,
\end{equation}
where $\mathbf{b}$ and $\beta$ are defined in (\ref{b1})--(\ref{b2}), and
$0\leq C_{q}\leq1$, $0\leq C^{\left(  \mu\right)  }\leq1$. The corresponding
\textquotedblleft scalar obscure concurrence\textquotedblright\ can be defined
as%
\begin{equation}
C_{scal}=\sqrt{\frac{\left\vert \det\mathbf{b}\right\vert ^{2}+\left\vert
\det\beta\right\vert ^{2}}{2}},
\end{equation}
such that $0\leq C_{scal}\leq1$. Thus, two obscure qubits are totally
separable, if $C_{scal}=0$.

For instance, for an obscure analog of the (maximally entangled) Bell state%
\begin{equation}
\left\vert \mathbf{\Psi}_{ob}\left(  2\right)  \right\rangle =\frac{1}%
{\sqrt{2}}\left(  \left[
\begin{array}
[c]{c}%
\frac{1}{\sqrt{2}}\\[5pt]%
\frac{1}{\sqrt{2}}%
\end{array}
\right]  \left\vert \mathsf{00}^{\prime}\right\rangle +\left[
\begin{array}
[c]{c}%
\frac{1}{\sqrt{2}}\\[5pt]%
\frac{1}{\sqrt{2}}%
\end{array}
\right]  \left\vert \mathsf{11}^{\prime}\right\rangle \right)
\end{equation}
we obtain%
\begin{equation}
\mathsf{C}_{vect}=\left[
\begin{array}
[c]{c}%
1\\
1
\end{array}
\right]  ,\ \ \ \ C_{scal}=1.
\end{equation}

A more interesting example is the \textquotedblleft intermediately
entangled\textquotedblright\ two obscure qubit state, e.g.%
\begin{equation}
\left\vert \mathbf{\Psi}_{ob}\left(  2\right)  \right\rangle =\frac{1}%
{\sqrt{2}}\left(  \left[
\begin{array}
[c]{c}%
\frac{1}{2}\\[5pt]%
\frac{1}{\sqrt{2}}%
\end{array}
\right]  \left\vert \mathsf{00}^{\prime}\right\rangle +\left[
\begin{array}
[c]{c}%
\frac{1}{4}\\[5pt]%
\frac{\sqrt{5}}{4}%
\end{array}
\right]  \left\vert \mathsf{10}^{\prime}\right\rangle +\left[
\begin{array}
[c]{c}%
\frac{\sqrt{3}}{4}\\[5pt]%
\frac{1}{2\sqrt{2}}%
\end{array}
\right]  \left\vert \mathsf{01}^{\prime}\right\rangle +\left[
\begin{array}
[c]{c}%
\frac{1}{\sqrt{2}}\\[5pt]%
\frac{1}{4}%
\end{array}
\right]  \left\vert \mathsf{11}^{\prime}\right\rangle \right)  , \label{ps22}%
\end{equation}
where the amplitudes satisfy (\ref{bb1})--(\ref{bb2}). If the Born-like rule
(as in (\ref{mm1})) holds for the membership amplitudes, then the
probabilities and membership functions of the states in (\ref{ps22}) are%
\begin{align}
p_{00^{\prime}}  &  =\frac{1}{4},\ \ p_{10^{\prime}}=\frac{1}{16}%
,\ \ p_{01^{\prime}}=\frac{3}{16},\ \ p_{11^{\prime}}=\frac{1}{2},\\
\mu_{00^{\prime}}  &  =\frac{1}{2},\ \ \mu_{10^{\prime}}=\frac{5}{16}%
,\ \ \mu_{01^{\prime}}=\frac{1}{8},\ \ \mu_{11^{\prime}}=\frac{1}{16}.
\end{align}
This means that, e.g., the state $\left\vert \mathsf{10}^{\prime}\right\rangle
$ will be measured with the quantum probability $1/16$ and the membership
function (\textquotedblleft truth\textquotedblright\ value) $5/16$. For the
entangled obscure qubit (\ref{ps22}) we obtain the concurrences%
\begin{equation}
\mathsf{C}_{vect}=\left[
\begin{array}
[c]{c}%
\frac{1}{2}\sqrt{2}-\frac{1}{8}\sqrt{3}\\[3pt]%
\frac{1}{8}\sqrt{2}\sqrt{5}-\frac{1}{4}\sqrt{2}%
\end{array}
\right]  =\left[
\begin{array}
[c]{c}%
0.491\\
0.042
\end{array}
\right]  ,\ \ \ C_{scal}=\sqrt{\frac{53}{128}-\frac{1}{16}\sqrt{5}-\frac
{1}{16}\sqrt{2}\sqrt{3}}=0.348.
\end{equation}

In the vector representation (\ref{akc})--(\ref{ps3}) we have%
\begin{equation}
\left\vert \mathsf{ij}^{\prime}\right\rangle =\left\vert \mathsf{i}%
\right\rangle \otimes\left\vert \mathsf{j}^{\prime}\right\rangle =\left[
\begin{array}
[c]{c}%
\mathsf{e}_{i}\otimes_{K}\mathsf{e}_{j^{\prime}}\\
\varepsilon_{i}\otimes_{K}\varepsilon_{j^{\prime}}%
\end{array}
\right]  ,\ \ \ \ i,j=0,1,\ \ \ j^{\prime}=0^{\prime},1^{\prime},
\end{equation}
where $\otimes_{K}$ is the Kronecker product (\ref{aa}), and $\mathsf{e}%
_{i},\varepsilon_{i}$ are defined in (\ref{e1})--(\ref{e2}). Using (\ref{bb})
and the Kronecker-like product (\ref{akc}), we put (informally, with no
summation)%
\begin{equation}
\mathsf{B}_{ij^{\prime}}\left\vert \mathsf{ij}^{\prime}\right\rangle =\left[
\begin{array}
[c]{c}%
b_{ij^{\prime}}\mathsf{e}_{i}\otimes_{K}\mathsf{e}_{j^{\prime}}\\
\beta_{ij^{\prime}}\varepsilon_{i}\otimes_{K}\varepsilon_{j^{\prime}}%
\end{array}
\right]  ,\ \ \ \ i,j=0,1,\ \ \ j^{\prime}=0^{\prime},1^{\prime}. \label{bij}%
\end{equation}

To clarify our model, we show here a manifest form of the two obscure qubit
state (\ref{ps22}) in the vector representation%
{\tiny
\begin{equation}
\left\vert \mathbf{\Psi}_{ob}\left(  2\right)  \right\rangle =\frac{1}%
{\sqrt{2}}\left(  \left[
\begin{array}
[c]{c}%
\frac{1}{2}\left(
\begin{array}
[c]{c}%
1\\
0\\
1\\
0
\end{array}
\right) \\[5pt]%
\frac{1}{\sqrt{2}}\left(
\begin{array}
[c]{c}%
1\\
0\\
1\\
0
\end{array}
\right)  ^{\left(  \mu\right)  }%
\end{array}
\right]  +\left[
\begin{array}
[c]{c}%
\frac{1}{4}\left(
\begin{array}
[c]{c}%
0\\
1\\
1\\
0
\end{array}
\right) \\[5pt]%
\frac{\sqrt{5}}{4}\left(
\begin{array}
[c]{c}%
0\\
1\\
1\\
0
\end{array}
\right)  ^{\left(  \mu\right)  }%
\end{array}
\right]  +\left[
\begin{array}
[c]{c}%
\frac{\sqrt{3}}{4}\left(
\begin{array}
[c]{c}%
1\\
0\\
0\\
1
\end{array}
\right) \\[5pt]%
\frac{1}{2\sqrt{2}}\left(
\begin{array}
[c]{c}%
1\\
0\\
0\\
1
\end{array}
\right)  ^{\left(  \mu\right)  }%
\end{array}
\right]  +\left[
\begin{array}
[c]{c}%
\frac{1}{\sqrt{2}}\left(
\begin{array}
[c]{c}%
0\\
1\\
0\\
1
\end{array}
\right) \\[5pt]%
\frac{1}{4}\left(
\begin{array}
[c]{c}%
0\\
1\\
0\\
1
\end{array}
\right)  ^{\left(  \mu\right)  }%
\end{array}
\right]  \right)  . \label{ps11}%
\end{equation}
}

The states above may be called \textquotedblleft symmetric two obscure qubit
states\textquotedblright. However, there are more general possibilities, as may be seen from the r.h.s. of (\ref{bij}) and (\ref{ps11}), when the indices of the
first and second rows do not coincide. This would allow more possible
states, which we call \textquotedblleft non-symmetric two obscure qubit
states\textquotedblright. It would be worthwhile to establish their
possible physical interpretation.

The above constructions show that quantum computing using
Kronecker obscure qubits can involve a rich structure of states, giving a more
detailed description with additional variables reflecting vagueness.

\section{Conclusions}

We have proposed a new scheme for describing quantum computation bringing vagueness into consideration, in which each state is characterized by a \textquotedblleft measure of truth\textquotedblright\. A
membership amplitude is introduced in addition to the probability amplitude in order to achieve this, and we are led thereby to the concept
of an obscure qubit. Two kinds of these are considered: the \textquotedblleft
product\textquotedblright\ obscure qubit, in which the total amplitude is the
product of the quantum and membership amplitudes, and the
\textquotedblleft Kronecker\textquotedblright\ obscure qubit, where the
amplitudes are manipulated separately. In latter case, the quantum part of the
computation is based, as usual, in Hilbert space, while the \textquotedblleft
truth\textquotedblright\ part requires a vague/fuzzy set formalism, and this can be
performed in the framework of a corresponding fuzzy space. Obscure-quantum computation may be considered as a set of rules (defining obscure-quantum gates) for managing quantum and membership amplitudes independently in different spaces. In this framework we obtain not only the probabilities of final
states, but also their membership functions, i.e. how much \textquotedblleft
trust\textquotedblright\ we should assign to these probabilities.
Our approach considerably extends the theory of quantum computing by adding the logic part directly to the computation process. Future challenges could lie in the direction of development of the corresponding logic hardware in parallel with the quantum devices.

\medskip

\textbf{Acknowledgments}. We acknowledge support from the Open Access Publication Fund of the University of M\"unster, Germany.

The first author (S.D.) is deeply thankful to Geoffrey Hare and Mike Hewitt for thorough language checking.


\end{document}